\documentclass[prl,showpacs,showkeys,nofootinbib,twocolumn,floatfix]{revtex4}
\usepackage{graphicx}  %
\usepackage{amsmath}
\usepackage{bm}  %

\newcommand{\bea}{\begin{eqnarray}}
\newcommand{\eea}{\end{eqnarray}}
\newcommand{\beq}{\begin{equation}}
\newcommand{\eeq}{\end{equation}}
\newcommand{\bqa}{\begin{eqnarray}}
\newcommand{\eqa}{\end{eqnarray}}

\def\mqo2{{\!\!\!}}

\begin{document}

\title{
Exact Relations for a Strongly-interacting Fermi Gas \\ 
from the Operator Product Expansion}

\author{Eric Braaten}\email{braaten@mps.ohio-state.edu}
\affiliation{Department of Physics,
         The Ohio State University, Columbus, OH\ 43210, USA\\}

\author{Lucas Platter}\email{lplatter@mps.ohio-state.edu}
\affiliation{Department of Physics,
         The Ohio State University, Columbus, OH\ 43210, USA\\}
\date{\today}
%\date{November 2007}

\begin{abstract}
The momentum distribution 
in a Fermi gas with two spin states and a large 
scattering length has a tail that falls off like  
$1/k^4$ at large momentum $k$, as pointed out by Shina Tan.
He used novel methods to derive exact relations between 
the coefficient of the tail in the momentum distribution
and various other properties of the system.
We present simple derivations of these relations 
using the operator product expansion for quantum fields.
We identify the coefficient as the integral over space
of the expectation value of a local operator that measures
the density of pairs.
\end{abstract}

\smallskip
\maketitle

Many-body systems of fermions have long been 
of great importance in astrophysics, nuclear physics, 
and solid state physics.
The development of trapping and cooling techniques for 
ultracold atoms has made them important in atomic physics as well.
In this case, the strength of the interaction is governed 
by the 2-body scattering length which can be controlled 
experimentally, adding a new dimension to the problem \cite{sps0706}.

If the scattering length $a$ is much larger than the range 
of the interactions, the system has universal properties 
that are determined only by the large scattering length.
For sufficiently low number density $n$, the universal
properties can be calculated using perturbative methods. 
If $n |a|^3$ is comparable to 1 or larger, 
the problem becomes nonperturbative.
In the special case of two equally-populated spin states,
systematically improvable
calculations are possible using Monte Carlo methods.
If the populations are not equal, this approach suffers
from the fermion sign problem.
If there are 3 or more spin states, the problem is complicated 
by the Efimov effect \cite{Braaten:2004rn}.
The challenging nature of the general problem makes exact results 
very valuable.  One case in which exact results are known 
is the unitary limit $a = \pm \infty$, where they can be derived
by exploiting scale invariance \cite{Ho04} and conformal
invariance \cite{Son:2005rv}.

In 2005, Shina Tan pointed out that the momentum distribution in
an arbitrary system consisting of fermions 
in two spin states with a large scattering length
has a large-momentum tail that falls off as
$1/k^4$ \cite{Tan0505}. 
The number of fermions with momentum larger than $K$ approaches
$C/(\pi^2 K)$ as $K \to \infty$,
where $C$ depends on the state of the system.  Tan used novel methods 
involving generalized functions to derive 
exact relations between $C$ and several 
other properties of the system.  An example is the
{\it adiabatic relation} that gives the change in the 
total energy $E$ due to a small change in $a$ \cite{Tan0508}:
%-----------------
\begin{equation}
\frac {d E \ \ }{d(1/a)} = - \frac{\hbar^2}{4\pi m} ~ C~.
\label{E-I}
\end{equation}
%-----------------
Tan referred to $C$ as the {\it integrated contact intensity},
which we will abbreviate to {\it contact}.
The Tan relations hold for any state 
of the system: few-body or many-body, homogeneous or in a
trapping potential, superfluid or normal, zero or
nonzero temperature.

In this Letter, we show that the Tan relations can be derived 
using the {\it operator product expansion} (OPE) for quantum fields. 
The OPE was proposed by Ken Wilson in 1969 
\cite{Wilson:1969} as a formalism for dealing with the strong 
interactions associated with the nuclear force. 
The OPE has become a standard tool to
understand the relativistic quantum field theories 
that describe elementary particles \cite{Collins}.  
We will apply the OPE to the 
strongly-interacting nonrelativistic system 
consisting of fermions with two spin states and
a large scattering length. 
We identify the contact $C$ as the expectation value 
of the integral over space of a local operator
that measures the density of pairs.
The OPE provides new insights into the Tan relations
and makes it easier to generalize them to systems with additional 
degrees of freedom or more complicated interactions.

A system consisting of fermions in two spin states labelled by 
$\sigma = 1, 2$ can be described by a quantum field theory 
with two quantum fields $\psi_\sigma(\bm{r})$. 
The number operator is
$\sum_\sigma \int d^3 R ~ \psi_\sigma^\dagger \psi_\sigma (\bm{R})$
\cite{footnote1}.
The momentum distribution $\rho_{\sigma}(\bm{k})$ for fermions 
with spin $\sigma$ can be expressed as
%-----------------
\begin{equation}
\rho_{\sigma} (\bm{k}) = 
\! \int \!\! d^3 R \!\! \int \!\! d^3r \;  e^{i \bm{k} \cdot \bm{r}} 
\langle \psi_\sigma^\dagger(\bm{R} \mbox{$-\frac12$} \bm{r})
\psi_\sigma(\bm{R} \mbox{$+\frac12$} \bm{r}) \rangle~.
\label{rho-psi}
\end{equation}
%-----------------
Its behavior at large $\bm{k}$ is determined by the matrix element 
at small $\bm{r}$. If the fermions are non-interacting, 
the quantum fields can be
expanded as Taylor series in $\bm{r}$. The resulting expansion for
$\rho_{\sigma}(\bm{k})$ can be expressed in terms of the Dirac delta function
in $\bm{k}$ and derivatives of the delta function. This indicates that
$\rho_{\sigma}(\bm{k})$ has a finite range in
$\bm{k}$. Simple examples are an ideal gas of fermions at 0 temperature, 
for which $\rho_{\sigma} (\bm{k})$ 
vanishes if $|\bm{k}|$ is larger than the Fermi momentum,
and an ideal gas of fermions at high temperature, for which 
$\rho_{\sigma}(\bm{k})$ is a Gaussian function of $\bm{k}$.

If there are interactions between the fermions, the matrix element in
Eq.~(\ref{rho-psi}) may not be an analytic function of $\bm{r}$
at $\bm{r}=0$.  The momentum distribution $\rho_{\sigma}(\bm{k})$ 
may therefore have a
large-momentum tail that falls off like a power of $k = |\bm{k}|$. 
Tan showed that for fermions with two spin states
and a large  scattering length, the tail is proportional to $1/k^4$
and is the same for both spin states: 
%-----------------
\begin{equation}
\rho_{\sigma} (\bm {k}) \longrightarrow C/k^4~,
\label{rho-I}
\end{equation}
%-----------------
where $C$ is the contact.
This power-law behavior can arise
from a term in the matrix element in Eq.~(\ref{rho-psi}) 
that is linear in $r = | \bm{r} |$ as $\bm{r} \to 0$. 
It holds for all states
that include at least one particle of each spin.
The dependence on the state enters only through $C$.

The possibility of a power-law tail in the momentum distribution can be
understood from Wilson's OPE.
Wilson proposed that in a quantum field theory a product of local operators 
separated by a short distance can be expanded in terms of local operators.
The expansion of the product of quantum fields in
Eq.~(\ref{rho-psi}) has the form
%-----------------
\begin{equation}
\psi_\sigma^\dagger(\bm{R} \mbox{$-\frac12$} \bm{r}) 
\psi_\sigma(\bm{R} \mbox{$+\frac12$} \bm{r}) = 
\sum_n C_{\sigma,n}(\bm{r}) {\cal O}_n (\bm{R})~,
\label{OPE}
\end{equation}
%-----------------
where the sum is over local operators ${\cal O}_n (\bm{R})$ that can be
constructed out of the quantum fields and their gradients. 
They include the operators that arise from multiplying  
the Taylor expansions of the operators on the left side 
of the equation, but there can be other operators as well.
The functions $C_{\sigma,n}(\bm{r})$ are called 
{\it Wilson coefficients} or {\it short-distance coefficients}.
Some of them can be nonanalytic at $\bm{r} = 0$, so
the momentum distribution $\rho_{\sigma}(\bm{k})$
can have a power-law tail.

%%%%%%%%%%%%%%%%%%%%%%%%%%%%%%%%%%%%%%%%%%%%%%%%%%
\begin{figure}[t]
\centerline{\includegraphics*[height=1.5cm,angle=0,clip=true]{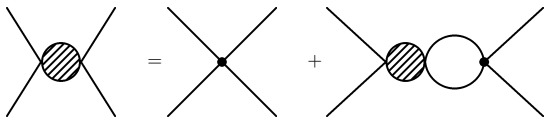}}
\vspace*{0.0cm}
\caption{(Color online) Integral equation for the scattering of a 
pair of fermions.  The blob represents the amplitude $i{\cal A}(E)$.
The solid dot represents the interaction vertex $-i g(\Lambda)$.
}
\label{fig:inteq}
\end{figure}
%%%%%%%%%%%%%%%%%%%%%%%%%%%%%%%%%%%%%%%%%%%%%%%%%%

The Hamiltonian density for the quantum field theory that describes 
fermions with two spin states and a large scattering length 
in an external potential $V(\bm{R})$ is
%-----------------
\begin{equation}
{\cal H} = 
\sum_\sigma \frac{1}{2m} 
        \nabla \psi_\sigma^\dagger \cdot \nabla \psi_\sigma^{(\Lambda)}
+ \frac{g(\Lambda)}{m} \psi_1^\dagger \psi_2^\dagger \psi_1 \psi_2^{(\Lambda)}
+ {\cal V}~,
\label{H}
\end{equation}
%-----------------
where ${\cal V} = V(\bm{R}) \sum_\sigma \psi_\sigma^\dagger \psi_\sigma$.
For simplicity, we have set $\hbar = 1$.
The superscripts $(\Lambda)$ on the operators in Eq.~(\ref{H})
indicate that their matrix elements are ultraviolet divergent 
and an ultraviolet cutoff is required to make them well defined. 
For the ultraviolet cutoff, we impose an upper limit 
$|\bm{k}|<\Lambda$ on momentum integrals.
In the limit $\Lambda \to \infty$, 
the Hamiltonian density in Eq.~(\ref{H}) describes fermions with 
zero-range interactions and scattering length $a$ if we take the 
coupling constant to be
%-----------------
\begin{equation}
g(\Lambda) = \frac{4 \pi a}{1 - 2 a \Lambda/\pi}~.
\label{g2}
\end{equation}
%-----------------
The amplitude for the scattering of a pair of fermions can be calculated 
by solving the Lippmann-Schwinger integral equation, 
which is represented diagrammatically in Fig.~\ref{fig:inteq}. 
The solution ${\cal A}(E)$ depends on the total energy $E$ of the 
pair of fermions in the center-of-mass frame and not 
separately on their momenta. 
After substituting Eq.~(\ref{g2}) for $g(\Lambda)$, 
the solution in the limit $\Lambda \to \infty$ is
%-----------------
\begin{equation}
{\cal A}(E) = \frac{4 \pi/m}{-1/a + \sqrt{- m E - i \epsilon}}~.
\label{A-E}
\end{equation}
%-----------------
The T-matrix element for scattering of a
pair of fermions with momenta $+\bm{p}$ and $-\bm{p}$
is obtained by setting $E = p^2/m$.

%%%%%%%%%%%%%%%%%%%%%%%%%%%%%%%%%%%%%%%%%%%%%%%%%%
\begin{figure}[t]
\centerline{\includegraphics*[height=3.9cm,angle=0,clip=true]{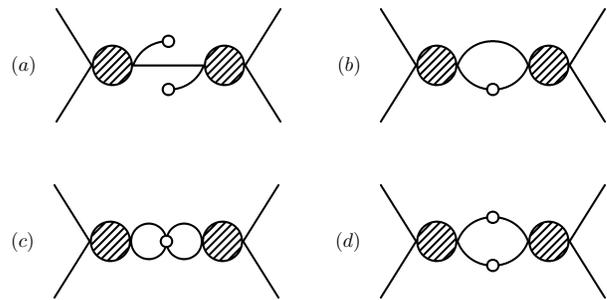}}
\vspace*{0.0cm}
\caption{(Color online) Diagrams for matrix elements of the 
operators (a) 
$\psi_\sigma^\dagger(-\frac12 \bm{r}) \psi_\sigma(+ \frac12 \bm{r})$,
(b) $\psi_\sigma^\dagger \psi_\sigma(\bm{0})$, (c)
$\psi_1^\dagger \psi_2^\dagger \psi_1 \psi_2^{(\Lambda)}(\bm{0})$, and (d)
$\psi_1^\dagger \psi_1(\mbox{$-\frac12$} \bm{r}) 
\psi_2^\dagger \psi_2(\mbox{$+\frac12$} \bm{r})$
between scattering states consisting of a pair of fermions.
The open dots represent the operators.
For each diagram, there are three other diagrams 
in which there is no scattering of the two incoming fermions 
or the two outgoing fermions or both.  }
\label{fig:<ops>}
\end{figure}
%%%%%%%%%%%%%%%%%%%%%%%%%%%%%%%%%%%%%%%%%%%%%%%%%%

We now proceed to show that the OPE in 
Eq.~(\ref{OPE}) includes a term with the operator 
${\cal O}_{12} = \psi_1^\dagger \psi_2^\dagger \psi_1 \psi_2^{(\Lambda)}$ 
and that its Wilson coefficient $C_{\sigma,12}(\bm{r})$ 
is linear in $r = |\bm{r}|$.  Since the OPE is an 
operator equation, $C_{\sigma,12}(\bm{r})$ can be determined by 
calculating the expectation value of both sides of the OPE in 
the simplest states for which 
$\langle\psi_1^\dagger \psi_2^\dagger \psi_1 \psi_2^{(\Lambda)} \rangle$
is nonzero.  We choose the state 
$| \pm \bm{p} \rangle$ consisting of two fermions 
with spins 1 and 2 and with momenta 
$+\bm{p}$ and $-\bm{p}$. 

We first consider the left side of the OPE in 
Eq.~(\ref{OPE}).  The operator product at $\bm{R} = 0$ can be 
represented diagrammatically by a pair of dots that correspond 
to the points $+ \frac12 \bm{r}$ where it annihilates an incoming fermion 
and $- \frac12 \bm{r}$ where it creates an outgoing fermion.  
The expectation value can be represented by the sum of the 
diagram in Fig.~\ref{fig:<ops>}(a) and the three diagrams 
with one or no scatterings.  
The contributions from the other three diagrams
are analytic at $\bm{r} = 0$. The contribution from the 
diagram in Fig.~\ref{fig:<ops>}(a) is
%-----------------
\begin{eqnarray}
\langle \psi_\sigma^\dagger(\mbox{$-\frac12$} \bm{r}) 
	\psi_\sigma(\mbox{$+\frac12$} \bm{r}) \rangle_{\pm \bm{p}} 
	\big|_{\rm \ref{fig:<ops>}(a)}
= im^2 {\cal A}^2(p^2/m) e^{i p r}/(8 \pi p)~.
\nonumber
\\
\label{<psipsi>}
\end{eqnarray}
%-----------------
If we expand this in powers of $r$, the terms with odd powers of $r$
are nonanalytic at $\bm{r}=0$.

We now consider the right side of the OPE in 
Eq.~(\ref{OPE}).  The expectation value of 
$\psi_\sigma^\dagger \psi_\sigma$
can be represented by the sum of the 
diagram in Fig.~\ref{fig:<ops>}(b) and the three diagrams 
with one or no scatterings.
The contribution from the diagram in 
Fig.~\ref{fig:<ops>}(b) is 
%-----------------
\begin{equation}
\langle \psi_\sigma^\dagger \psi_\sigma(\bm{0}) \rangle_{\pm \bm{p}} 
	\big|_{\rm \ref{fig:<ops>}(b)}
= im^2 {\cal A}^2(p^2/m)/(8 \pi p)~.
\label{<psi^2>}
\end{equation}
%-----------------
This matches the $r^0$ term in the expansion of 
Eq.~(\ref{<psipsi>}) in powers of $r$.  Thus we find that the 
Wilson coefficient of
$\psi_\sigma^\dagger \psi_\sigma(\bm{R})$ in Eq.~(\ref{OPE}) is simply 1
in accord with naive expectations.
To match the term linear in $r$ in the expansion of Eq.~(\ref{<psipsi>}), 
we must find an operator whose 
expectation value in the state $| \pm \bm{p} \rangle$
has the momentum dependence ${\cal A}^2(p^2/m)$.
One can deduce this operator by realizing that the nonanalytic behavior
at $\bm{r} = 0$ must arise from the region of the diagram in 
Fig.~\ref{fig:<ops>}(a) 
in which there is large momentum flowing in the lines that connect
the operators to the amplitudes and the line that connects the two amplitudes.
If we shrink all these lines to a point,
they reduce to a vertex with two incoming lines and two outgoing lines.
The simplest corresponding operator is 
${\cal O}_{12} = \psi_1^\dagger \psi_2^\dagger \psi_1 \psi_2^{(\Lambda)}$.
We can verify that this is the correct operator by calculating 
its matrix element, which can be represented by the sum of the 
diagram in Fig.~\ref{fig:<ops>}(c) and the three diagrams 
with one or no scatterings. 
By using the integral equation in Fig.~\ref{fig:inteq}, 
the sum of the four diagrams can be expressed in the simple form
%-----------------
\begin{equation}
\langle \psi_1^\dagger \psi_2^\dagger 
        \psi_1 \psi_2^{(\Lambda)} (\bm{0}) \rangle_{\pm \bm{p}}
= m^2 g^{-2}(\Lambda) {\cal A}^2(p^2/m)~.
\label{<V>}
\end{equation}
%-----------------
This has the same dependence on $p$ as the term linear in $r$
in the expansion of Eq.~(\ref{<psipsi>}).
To match this term, the Wilson coefficient must be
$C_{\sigma,12}(\bm{r}) = - g^2(\Lambda) \, r/(8 \pi)$.
The expectation value of this term in OPE can be written
%-----------------
\begin{equation}
C_{\sigma,12}(\bm{r}) \langle {\cal O}_{12}(\bm{R}) \rangle = 
- \frac{r}{8 \pi} 
\langle g^2 \psi_1^\dagger \psi_2^\dagger
              \psi_1 \psi_2 (\bm{R})  \rangle~.
\label{C12O12}
\end{equation}
%-----------------
We have attached the factor $g^2(\Lambda)$ to the operator
$\psi_1^\dagger \psi_2^\dagger \psi_1 \psi_2^{(\Lambda)}$
and suppressed the dependence on $\Lambda$,
because the resulting operator has finite matrix elements
in the limit $\Lambda \to \infty$,
as exemplified by Eq.~(\ref{<V>}).
We proceed to use this result to derive the Tan relations.

{\bf Tail of the momentum distribution}.
The leading behavior of $\rho_{\sigma}(\bm{k})$ at large $\bm{k}$ 
can be obtained by inserting the term in Eq.~(\ref{C12O12})
in place of the matrix element in Eq.~(\ref{rho-psi}).
This term gives the asymptotic behavior in 
Eq.~(\ref{rho-I}) with 
%-----------------
\begin{equation}
C = \int \!\! d^3R \, 
\langle g^2 \psi_1^\dagger \psi_2^\dagger 
                      \psi_1 \psi_2 (\bm{R}) \rangle ~.
\label{IX}
\end{equation}
%-----------------
We will refer to $g^2 \psi_1^\dagger \psi_2^\dagger \psi_1 \psi_2$
as the {\it contact density operator}.
This is a positive operator, so $C \ge 0$.

{\bf Energy relation}.
In Ref.~\cite{Tan0505}, Shina Tan derived an expression for the energy 
$E$ as a linear functional of the momentum distributions:
%-----------------
\begin{equation}
E = 
\sum_\sigma \int \!\! \frac{d^3k}{(2 \pi)^3} \frac{k^2}{2m} 
        \bigg( \rho_{\sigma}(\bm{k}) - \frac{C}{k^4} \bigg)
+ \frac{C}{4 \pi m a} 
+ \int \!\! d^3R \, \langle{\cal V}\rangle~.
\label{E-finite}
\end{equation}
%-----------------
This is a functional of $\rho_1$ and $\rho_2$, because 
$C$ is determined by their large-momentum behavior:
$C = \lim_{k \to \infty} k^4 \rho_\sigma(\bm{k})$.
Tan's energy relation can be derived simply by 
using the expression for $g(\Lambda)$ in Eq.~(\ref{g2})
to express the 
Hamiltonian density in Eq.~(\ref{H}) as the sum of three terms 
whose matrix elements are ultraviolet finite:
%-----------------
\begin{eqnarray}
{\cal H} = 
\Big( \sum_\sigma \frac{1}{2m} 
\nabla \psi_\sigma^\dagger \cdot \nabla \psi_\sigma^{(\Lambda)}
- \frac{m\Lambda}{2 \pi^2} 
g^2 \psi_1^\dagger \psi_2^\dagger \psi_1 \psi_2 \Big)
\nonumber
\\
+ \frac{m}{4 \pi a} 
g^2 \psi_1^\dagger \psi_2^\dagger \psi_1 \psi_2
+ {\cal V}~.
\label{H-finite}
\end{eqnarray}
%-----------------
The operator
$\nabla \psi_\sigma^\dagger \cdot \nabla \psi_\sigma^{(\Lambda)}$
in the first term has matrix elements that diverge linearly 
as $\Lambda \to \infty$.  This linear divergence is cancelled 
by the second term, which has an explicit factor of $\Lambda$.
Integrating over the positions
of the local operators in Eq.~(\ref{H-finite}), 
taking the expectation value, and using the  
expression for $C$ in Eq.~(\ref{IX}), we obtain
Tan's energy relation in Eq.~(\ref{E-finite}).

{\bf Adiabatic relation}.
Eq.~(\ref{E-I}) can be derived by using the Feynman-Hellman theorem:
%-----------------
\begin{equation}
d E/da = 
\int \!\! d^3R \, \langle \partial{\cal H}/\partial a \rangle ~.
\label{E-H}
\end{equation}
%-----------------
Since ${\cal H}$
depends on $a$ only through the coupling constant $g(\Lambda)$
in Eq.~(\ref{g2}), its derivative can be written
%-----------------
\begin{equation}
\partial{\cal H}/\partial a =
g^2 \psi_1^\dagger \psi_2^\dagger \psi_1 \psi_2/(4 \pi m a^2) ~.
\label{H-a}
\end{equation}
%-----------------
Upon inserting this into Eq.~(\ref{E-H})
and using Eq.~(\ref{IX}) for $C$,
we get the adiabatic relation in Eq.~(\ref{E-I}).

{\bf Virial theorem}.
The virial theorem for fermions with two spin states 
in a harmonic trapping potential in the unitary limit 
$a = \pm \infty$ was derived in Ref.~\cite{TKT0504}.
Tan has derived a generalization of the virial theorem 
for the case of finite $a$ \cite{Tan0803}.
The virial theorem
can be derived more simply by using the fact that the
scattering length $a$ and the angular frequency $\omega$ 
of the trapping potential
provide the only scales for the energy $E$ of a state. 
Dimensional analysis then requires the differential operator
$(\omega \partial/\partial \omega) 
- \frac12 (a \partial/\partial a)$ to give 1 when acting on
$\int \! d^3R \, \langle {\cal H} \rangle$.
Using the Feynman-Hellman theorem together with Eq.~(\ref{H-a}), 
we obtain the virial theorem 
%-----------------
\begin{equation}
E = 2 \int \!\! d^3R \, \langle{\cal V}\rangle - C/(8 \pi m a)~.
\label{virial}
\end{equation}
%-----------------

We can obtain a simple interpretation of the contact density operator
by considering the OPE of the number density 
operators $\psi_1^\dagger \psi_1(-\frac12 \bm{r})$ 
and $\psi_2^\dagger \psi_2(+\frac12 \bm{r})$.
The expectation value of their product in the state 
$| \pm \bm{p} \rangle$
can be represented by the sum of the 
diagram in Fig.~\ref{fig:<ops>}(d) and the three diagrams 
with one or no scatterings.  
The contribution that is most singular as $\bm{r} \to 0$
comes from the diagram in Fig.~\ref{fig:<ops>}(d):
$m^2 {\cal A}^2(p^2/m) e^{2 i p r}/(16 \pi^2 r^2)$.
The term proportional to $r^{-2}$
has the same dependence on $p$ as the matrix element in Eq.~(\ref{<V>}).
Thus the most singular term in the OPE is \cite{Tan0505}
%-----------------
\begin{eqnarray}
\psi_1^\dagger \psi_1(\bm{R} \mbox{$-\frac12$} \bm{r}) 
\psi_2^\dagger \psi_2(\bm{R} \mbox{$+\frac12$} \bm{r}) 
\to  \frac{1}{16 \pi^2 r^2}
g^2 \psi_1^\dagger \psi_2^\dagger \psi_1 \psi_2(\bm{R})~.
\nonumber
\\
\label{psi4-ope}
\end{eqnarray}
%-----------------
We can define an operator $N_{\rm pair}(\bm{R}, s)$
that counts the number of pairs of fermions with spins 1 and 2
near the point $\bm{R}$ with separation less than $s$
by integrating the left side of Eq.~(\ref{psi4-ope})
over the ball $|\bm{r}| < s$.  In the absence of interactions, 
$\langle N_{\rm pair}(\bm{R}, s) \rangle$ 
scales as $s^3$ as $s \to 0$.
Eq.~(\ref{psi4-ope}) implies that 
in the case of a large scattering length
$\langle N_{\rm pair}(\bm{R}, s) \rangle$ scales as $s^1$.
We can interpret the contact density operator
$g^2 \psi_1^\dagger \psi_2^\dagger \psi_1 \psi_2$
as the limit as $s \to 0$ of
$(4 \pi/s) N_{\rm pair}(\bm{R}, s)$.

One can give a more intuitive interpretation of the contact $C$
if the fermions with spins 1 and 2 have an inelastic 
two-body scattering channel into other spin states 
that are much lower in energy.  In this case,
the optical theorem implies that the scattering length $a$ 
has a negative imaginary part.  The leading effects of a 
weakly-coupled inelastic channel on low-energy fermions 
in the spin states of interest can be taken into account 
through the small imaginary part of $a$.
The effect on a state with definite energy $E$ 
is to change its time-dependence from
$\exp(-i E t/\hbar)$ to $\exp(-i (E - i \Gamma/2)t/\hbar)$.
The probability in that state decreases with time at the rate 
$\Gamma/\hbar$.  The adiabatic relation in Eq.~(\ref{E-I})
can be used to derive an expression for $\Gamma$ to 
leading order in the imaginary part of $a$:
%-----------------
\begin{equation}
\Gamma \approx \frac{\hbar^2 (-{\rm Im}\,a)}{2 \pi m |a|^2} C ~.
\label{Gam-I}
\end{equation}
%-----------------
Thus $C$ determines the rate at which 
low-energy fermions are depleted by inelastic collisions.

In Ref.~\cite{Tan0508}, Tan derived expressions for the contact 
density in the BCS limit ($a \to 0^-$), the unitary limit
($a \to \pm \infty$), and the BEC limit ($a \to 0^+$)
by using the adiabatic relation in Eq.~(\ref{E-I}) 
as an operational definition of $C$. 
Our identification of the local contact density operator
$g^2 \psi_1^\dagger \psi_2^\dagger \psi_1 \psi_2$
makes it straightforward to calculate the contact density 
for homogeneous systems directly using diagrammatic methods.

The Tan relations for ultracold fermionic atoms with two spin states
apply equally well to cold neutron matter at sufficiently low densities.
One advantage of our derivation using the OPE
is that it makes it easier to generalize the Tan relations to 
more complicated systems.  The generalization to 
more complicated interactions is straightforward for any system 
that can be described by a renormalizable 
local quantum field theory \cite{BKZ0709}.
An important example is the resonance model that provides 
a natural description of atoms near a Feshbach resonance \cite{KMCWH02}.
The generalization to systems with additional spin degrees of freedom,
such as nuclear matter at sufficiently low densities, is
complicated by the Efimov effect \cite{Braaten:2004rn}.
In such cases, a 3-body analog of the contact density operator 
may be expected to play an important role.

The original definition of the contact $C$ in terms of the 
tail of the momentum distribution suggests that the Tan relations
are relevant only to esoteric aspects of the strongly-interacting 
Fermi gas.  However the adiabatic relation in Eq.~(\ref{E-I})
and the virial theorem in Eq.~(\ref{virial}) make it clear that they
are actually of central importance.
In Refs.~\cite{Tan0505,Tan0508}, Tan offered suggestions for how
these relations could be tested experimentally.
We leave this as a challenge to the ingenuity of experimentalists 
in cold atom physics.
 
\begin{acknowledgments}
We thank Shina Tan for valuable discussions.
This research was supported in part by the Department of Energy 
under grants DE-FG02-05ER15715 and DE-FC02-07ER41457 and
by the National Science Foundation under grant PHY-0653312.
\end{acknowledgments}


\begin{thebibliography}{99}
\bibitem{footnote1}
We denote the product of local
operators ${\cal O}_A (\bm{r})$ and ${\cal O}_B (\bm{r})$ at the same
point in space by ${\cal O}_A {\cal O}_B (\bm{r})$.
\end{thebibliography}
\end{document}